\documentclass{article}
\usepackage[numbers]{natbib}
\usepackage[preprint]{neurips_2023}
\usepackage[utf8]{inputenc} 
\usepackage[T1]{fontenc}    
\usepackage{hyperref}       
\usepackage{url}            
\usepackage{booktabs}       
\usepackage{amsfonts}       
\usepackage{amsmath}
\usepackage{makecell}
\usepackage{microtype}      
\usepackage{verbatim}
\usepackage{courier}
\usepackage{float}
	
\makeatletter
\renewcommand{\@notice}{}
\makeatother

\title{WebApp1K: A Practical Code-Generation Benchmark for Web App Development}
\author{%
	Yi Cui \\
	ONEKQ Lab\\
	{yi@onekq.ai} \\
}

\begin{document}	
\maketitle

\begin{abstract}
We introduce WebApp1K, a practical code-generation benchmark to measure LLM ability to develop web apps. This benchmark aims to calibrate LLM output and aid the models to progressively improve code correctness and functionality. The benchmark is lightweight and easy to run.

We present the initial version of WebApp1K, and share our findings of running the benchmark against the latest frontier LLMs. First, open source LLMs deliver impressive performance, closely trailing behind GPT-4o and Claude 3.5. Second, model size has strong correlation with code correctness. Third, no prompting techniques have been found to lift performance either universally to all models, or significantly to a single model.
\end{abstract}
	
\section{Introduction}
Large language models (LLM) have been widely adopted in the software industry to enhance project velocity and code quality. Many coding benchmarks have been developed focusing on different aspects such as algorithms\cite{mbpp}, data science\cite{ds1000}, class-level code\cite{classeval}, execution\cite{codereval}, resolving software bugs\cite{swebench}, etc., more efforts are needed to calibrate LLMs' ability to assist software development for consumer-grade applications. While the current norm of LLM application is in the form of copilot or code assistant to existing software developers\cite{survey}, we argue that LLMs already, and will continue to, empower more and more non-developers (by traditional standards) to write code. Below are some scenarios.

\begin{itemize}
\item A handful of first-time startup builders brainstorm a new social network helping a group of underprivileged individuals, and bootstrap themselves to launch a prototype to the public. LLMs play pivotal role during each phase of the project, e.g. planning, tech stack choice, task assignment, code generation, and deployment.

\item A small group of employees volunteer to a greenfield project funded by a mid-sized software company. While enjoying the freedom of scope and high-level blessing, the team finds itself understaffed and under-resourced. The founding members spares no time to reach an all-hands-on-deck decision, and turns everyone into a developer with the aid of LLMs, very similar to the above startup scenario.

\item A college student in their senior year, determined to be a top competitor on the job market, decided to boost their GitHub profile by some impressive demos. They first consulted LLMs on popular frameworks and best practices, then delved into a project and continue to utilize LLMs to help maximize velocity. In particular, LLMs are used to generate, from scratch, functional code expected to be bug-free and ready-to-build.
\end{itemize}

Arguably, the ensemble of users exemplified above is a much larger group than the professional developer community. Undoubtedly LLMs will deliver tremendous and life-altering values to this group of users. However, as will be revealed by this study, the improvement space for LLMs is huge. To this end, we present a preliminary work to benchmark LLM performance on practical application development.

We aim the consumers of this benchmark to be practitioners working closely with LLMs, e.g. those who ship models (pre-training and post-training), or those who build LLM applications and decide how to choose/use their models. Lots of considerations are taken into the benchmark usability.

\begin{itemize}
	\item \textbf{\textit{Marketability}} Since the benchmark aims to evaluate usefulness of LLMs to practical software development, the model artifact, i.e. the generated code, must meet the needs of customers or employers. As such, LeetCode-type algorithmic problems, while extremely valuable in other contexts, are not applicable here. Moreover, the evaluation would be a lot easier if the knowledge and capabilities to create marketable code is inherently possessed by LLMs, i.e. present in their training dataset.

	\item \textbf{\textit{Metric}} In order to present a clear direction to model hill climbing, the metric should be simple with unambiguous answer to each problem of the benchmark. Fortunately, low-hanging fruits are available because the code correctness as the most basic criteria is still an unsolved problem.

	\item \textbf{\textit{Overhead}} Training LLMs is a high-budget and high-risk process introducing lots of stress. To exacerbate the problem, training environments and tools are primitive, also vulnerable to deprecation given the continuously escalating model scale and the training infra it demands. Therefore, one should assume that bare-bone Linux is all we can count on. Anything more sophisticated, e.g. sandbox or virtual environment, will decrease the benchmark adopt-ability, also taking more time to run, potentially causing training delays.
\end{itemize}

We are blessed by achievements of open source contributors to address all these challenges. By building our benchmark around the most successful open-source frameworks, the generated code is widely considered marketable by the industry. Also these frameworks have abundant quality code and best practices shared on the Internet, already handily included in the training dataset of modern LLMs. Finally, the rich ecosystem of these frameworks equips us with stable and well-distributed testing tools to address the metric and overhead challenges.

We introduce WebApp1K, a benchmark to evaluate web app code generation performance of LLMs. We start with React, a JavaScript framework widely recognized as the top choice for web app development. Below are our main findings:

\begin{enumerate}
	\item Open source models deliver strong results. The top performing model is DeepSeek Coder V2 which closely trails behind GPT-4o and Claude 3.5 Sonnet, the top two LLMs.

	\item There is significant performance gap between the big model and the small model of the same family, proving that the scaling laws still apply in our evaluation context of code generation.
	
	\item No prompting techniques have been found to universally improve performance for all models or significantly lift performance of a single model. On the other hand, the strong performance of top performers (GPT and Claude) under a very simple prompt suggests that LLM should be able to code correctly as long as the expectations (i.e. test cases) are clear.
\end{enumerate}

We share three artifacts on GitHub and Huggingface: the dataset containing all 1000 problems of WebApp1K\cite{webapp1k-dataset}, the script\cite{webapp1k-script} to run WebApp1K, and the leaderboard\cite{webapp1k-leaderboard}.

The rest of this report is organized as follows. Section \ref{sec:benchmark} details how WebApp1K was built. Section \ref{sec:evaluation} presents our evaluation methodology, various experiments, and performance results of the latest LLMs. Finally, Sec \ref{sec:conclusion}, conclude contributions and discusses future works. 
\section{Benchmark}\label{sec:benchmark}
We choose React\cite{react} as the first evaluation subject of our benchmark. Introduced in 2013, it has remained a top choice for web app builders for the quality and stability of the framework itself, productivity gains to developers, and the rich ecosystem. In addition to these marketabiliy benefits, it also constructs a higher level abstraction above HTML syntax and DOM manipulation, allowing developers to encapsulate their business logic to succinct and reusable code. These technical advantages also benefit the code generation evaluation on LLMs.

Consider a blogging website, in which a user adds comment to an existing blog post. This user journey is simulated by the unit test in Table \ref{tab:success}. Here, \texttt{fetchMock.post} is a lightweight setup to mock a successful API response without running any additional software components. The following \texttt{await} lines simulate user actions (text input, mouse click etc.). Finally, \texttt{expect} lines examine the expected outcome, i.e. the mocked API should be invoked exactly once and the system response of success should appear on the updated webpage. Similarly, the pairing failure case is shown in Table \ref{tab:failure}, where a mocked API failure is expected to lead to error message on the updated webpage.
\begin{table}[!t]
	\centering
	\begin{tabular}{|l|}
		\hline
		\begin{minipage}{\dimexpr\textwidth-2\fboxsep-2\fboxrule}
			\vspace{2mm}
			\begingroup
			\renewcommand{\ttdefault}{pcr} 
			\scriptsize
			\begin{verbatim}
				test("successfully adds a comment to a post", async () => {
				  fetchMock.post("/api/comments", 200);

				  await act(async () => { 
				    render(<MemoryRouter><App /></MemoryRouter>); 
				  });
				  await act(async () => { 
				    fireEvent.change(screen.getByPlaceholderText(/Add a comment/i), 
				    { target: { value: "Great post!" } }); 
				  });
				  await act(async () => { 
				    fireEvent.click(screen.getByText(/Submit/i)); 
				  });

				  expect(fetchMock.calls("/api/comments").length).toBe(1);
				  expect(screen.getByText(/Comment added successfully/i)).toBeInTheDocument();
				}, 10000);
			\end{verbatim}
			\endgroup
		\end{minipage} \\
		\hline
	\end{tabular}
	\caption{Success Case for Adding a Comment to a Blog Post}
	\label{tab:success}
\end{table}

\begin{table}[!t]
	\centering
	\begin{tabular}{|l|}
		\hline
		\begin{minipage}{\dimexpr\textwidth-2\fboxsep-2\fboxrule}
			\vspace{2mm}
			\begingroup
			\renewcommand{\ttdefault}{pcr} 
			\scriptsize
			\begin{verbatim}
				test("fails to add a comment to a post", async () => {
				  fetchMock.post("/api/comments", 500);

				  // Lines identical to the success case are ignored.

				  expect(screen.getByText(/Failed to add comment/i)).toBeInTheDocument();
				}, 10000);
			\end{verbatim}
			\endgroup
		\end{minipage} \\
		\hline
	\end{tabular}
	\caption{Failure Case for Adding a Comment to a Blog Post}
	\label{tab:failure}
\end{table}

Table \ref{tab:impl} shows the verbatim code generated by GPT-4o as an attempt to pass these two tests, which it does. Besides the toyish API and purposely simplified application semantics, the code is functional and conceptually reusable to a web application which needs social commenting functionality.
\begin{table}[!t]
	\centering
	\begin{tabular}{|l|}
		\hline
		\begin{minipage}{\dimexpr\textwidth-2\fboxsep-2\fboxrule}
			\vspace{2mm}
			\begingroup
			\renewcommand{\ttdefault}{pcr} 
			\scriptsize
			\begin{verbatim}
			import React, { useState } from 'react';

			function App() {
			  const [comment, setComment] = useState('');
			  const [message, setMessage] = useState('');
	
  			  const handleSubmit = async (e) => {
			    e.preventDefault();
			    try {
			      const response = await fetch('/api/comments', {
			        method: 'POST',
			        headers: {
			          'Content-Type': 'application/json',
			        },
			        body: JSON.stringify({ comment }),
			      });
			
			      if (response.ok) {
			        setMessage('Comment added successfully');
			      } else {
			        throw new Error('Failed to add comment');
			      }
			    } catch (error) {
			      setMessage('Failed to add comment');
			    }
			  };
	
			  return (
			    <div>
			      <form onSubmit={handleSubmit}>
			        <input
			          type="text"
			          value={comment}
			          onChange={(e) => setComment(e.target.value)}
			          placeholder="Add a comment"
			        />
			        <button type="submit">Submit</button>
			      </form>
			      {message && <p>{message}</p>}
			    </div>
			  );
			}

			export default App;
			\end{verbatim}
			\endgroup
		\end{minipage} \\
		\hline
	\end{tabular}
	\caption{Code generated by GPT-4o to pass tests in Tables \ref{tab:success} and \ref{tab:failure} }
	\label{tab:impl}
\end{table}

The WebApp1K benchmark consists of 1000 problems, each depicting an unique user journey illustrated above. The construction of this benchmark is a combination of human curating and data synthesis by LLMs.

First, human proposes 20 web apps and their feature sets. They are listed in Table \ref{tab:scenarios} of the Appendix section. Then, within the scope of each app, we ask GPT-4o to describe 50 different user journeys, then review and iterate with the LLM until quality and relevancy are satisfactory. Finally, we feed each user journey to GPT-4o, prompting it to codify the journey into a success case and a failure case. Each of them strictly follows the \texttt{fetchMock-await-expect} sequence illustrated in Table \ref{tab:success}. All test cases can be found in our GitHub repo\cite{webapp1k-script}.
\section{Evaluation Results}\label{sec:evaluation}
\subsection{Experiment Setup}
The most straightforward way for us to access LLMs are public token-based APIs. For top close-sourced models, our only option is via the owners' APIs. The top open-sourced models are hosted by a few platforms, among which we choose Fireworks\cite{fireworks}.

We evaluate three closed-source, and three open-source model families listed in Table \ref{tab:models}.
\begin{table}[h]
	\caption{LLMs evaluated by WebApp1K}
	\centering
	\begin{tabular}{|l|c|c|c|}
		\hline
		& \textbf{Large} & \textbf{Small} & \textbf{Open/closed} \\
		\hline
		\textbf{GPT} & gpt-4o\cite{gpt-4o} & gpt-4o-mini\cite{gpt-4o-mini} & Closed \\
		\hline
		\textbf{Claude} & claude-3.5-sonnet\cite{claude} & N/A\footnote{We have considered Claude Haiku but dropped because it belongs to v3, not v3.5} & Closed \\
		\hline
		\textbf{Gemini} & gemini-1.5-pro\cite{gemini-1.5-pro} & gemini-1.5-flash\cite{gemini-1.5-flash} & Closed \\
		\hline
		\textbf{DeepSeek Coder V2} & deepseek-coder-v2-instruct\cite{deepseek} & deepseek-coder-v2-lite-instruct\cite{gpt-4o-mini} & Open \\
		\hline
		\textbf{Llama 3} & llama-v3-70b-instruct\cite{llama3-70b} & llama-v3-8b-instruct\cite{llama3-8b} & Open \\
		\hline
		\textbf{Mixtral} & mixtral-8x22b-instruct\cite{mixtral-8x22b} & mixtral-8x7b-instruct\cite{mixtral-8x7b} & Open \\
		\hline
	\end{tabular}
	\label{tab:models}
\end{table}

Below is the prompt we use. The majority of it is the test code exemplified in Tables \ref{tab:success} and \ref{tab:failure}.
\begin{align}
	&\text{Generate }\{file\_name\}\text{ to pass the tests below: } \label{eq:prompt} &\\
	&\{success\_test\_code\}\{failure\_test\_code\}.\text{ RETURN CODE ONLY.} \nonumber &
\end{align}

We use $pass@k$, a metric defined in \cite{humaneval} and commonly accepted by subsequent coding task evaluation works. Due to budget and rate limit constraints, we evaluate each LLM-problem pair 10 times, i.e. $n=10$. Since $k$ must be smaller or equal to $n$, we measure $pass@1$, $pass@5$, and $pass@10$. 
\subsection{Parameter Tuning}
Although each API bears its minor difference, all APIs are heavily influenced by the design of OpenAI API. Table \ref{tab:api_params} lists the tunable parameters exposed by each API. Since we do not know the default parameter value set by each API provider, we explicitly set the same parameter values to all LLMs under evaluation, whenever applicable. To limit the search space, we only tune $temperature$ and $top\_p$, the two most popular parameters available on all platforms. For other parameters, we assign fixed value to them across all LLMs.
\begin{table}[h]
	\caption{Tunable Parameters on Different APIs}
	\centering
	\begin{tabular}{|l|c|c|c|c|c|}
		\hline
		& \textbf{temperature} & \textbf{top\_p} & \textbf{top\_k} & \textbf{presence\_penalty} & \textbf{frequency\_penalty} \\
		\hline
		\textbf{GPT4o} & Y & Y & N & Y & Y \\
		\hline
		\textbf{Claude} & Y & Y & Y & N & N \\
		\hline
		\textbf{Gemini} & Y & Y & Y & N & N \\
		\hline
		\textbf{Fireworks} & Y & Y & Y & Y & Y \\
		\hline
	\end{tabular}
	\label{tab:api_params}
\end{table}

We conducted a grid search to locate a sweet spot at which all LLMs deliver near-best results. We chose 100 random problems from the benchmark, 5 out of each application. We then choose the large model out of the five leading model families, and measured their $pass@1$ ($n=1$) on the discrete 2D space of $temperature$ and $top\_n$, where $temperature = 0, 0.1, 0.2, ..., 1$, and $top\_p = 0, 0.1, 0.2, ..., 1$.
\begin{table}[h]
	\caption{Parameter Tuning Results on $Pass@1$}
	\centering
	\begin{tabular}{|l|c|c|c|}
		\hline
        & \textbf{Lowest} & \makecell{\textbf{Chosen} \\ ($\text{temperature} = 0.2$, $\text{top\_p} = 0.8$)} & \textbf{Highest} \\
		\hline
		gpt-4o & 0.81 & \textbf{0.88} & 0.9 \\
		claude-3.5-sonnet & 0.82 & \textbf{0.85} & 0.86 \\
		deepsseek-coder-v2-instruct & 0.42 & \textbf{0.59} & 0.59 \\
		gemini-1.5-pro & 0.59 & \textbf{0.65} & 0.69 \\
		llama-v3-70b-instruct & 0.19 & \textbf{0.31} & 0.34 \\
		\hline
	\end{tabular}
	\label{tab:optimal_params}
\end{table}

Table \ref{tab:optimal_params} presents the lowest and highest $Pass@1$ value by each LLM in this grid search. Based on the results, we finalize our parameters as follows.
\begin{align*}
	temperature = 0.2 \\
	top\_p = 0.8 \\
	top\_k = 40 \\
	presence\_penalty = 0 \\
	frequency\_penalty = 0
\end{align*}

As will be shown later, results of our full-scale evaluations also align with this small-scale experiment, except for the Deepseek coder model\cite{deepseek} whose performance exceeds expectation. Also worth noting is that open-source models exhibit larger performance variation than closed-source models.
\subsection{Prompt Experiments}
We also study whether more sophisticated can lift the model performance.

The first experiment is \textbf{\textit{system prompt}}, which assigns an explicit role to the LLM and raises its awareness. Available in all APIs we run, it complements the user prompt (Equation (\ref{eq:prompt})) which gives detailed instructions to LLM. Equation (\ref{eq:system_prompt}) shows our system prompt.
\begin{equation}\label{eq:system_prompt}
	\text{You are a code generator.}
\end{equation}

The second experiment is \textbf{\textit{verbose comment}}, which aims to help LLMs better understand the semantics of  tests it tries to pass. For each of the 1000 problems, we feed its test code to GPT-4o and ask for English summary of the expectation in multiple sentences. The summary is then inserted into the test code. Table \ref{tab:verbose} shows the verbose comment variant of the test code in Table \ref{tab:success}.

\begin{table}[!t]
	\centering
	\begin{tabular}{|l|}
		\hline
		\begin{minipage}{\dimexpr\textwidth-2\fboxsep-2\fboxrule}
			\vspace{2mm}
			\begingroup
			\renewcommand{\ttdefault}{pcr} 
			\scriptsize
			\begin{verbatim}
				test(
				"This test case verifies that a comment can be successfully added to a post by simulating
				a successful POST request to the '/api/comments' endpoint. The test ensures that the
				API call occurs exactly once and that a success message ('Comment added successfully')
				is displayed upon successful submission. This helps confirm the correct interaction
				between the frontend and backend components when adding comments.",
				  async () => {

				  // Lines identical to the original test case are ignored.

				}, 10000);
			\end{verbatim}
			\endgroup
		\end{minipage} \\
		\hline
	\end{tabular}
	\caption{Verbose Comment Variant of the Test Case in Table \ref{tab:success}}
	\label{tab:verbose}
\end{table}
 
The third experiment is \textbf{\textit{error debugging}}. If the generated code fails the test, we add the failed code and the error log to the prompt, hoping the LLM will generate the correct code by learning from its own mistakes. Below is the prompt.
\begin{align*}
	&\{failed\_implementation\} \\
	&\text{The above code is the implementation of }\{file\_name\}\text{. It failed the tests below}\\
	&\{success\_test\_code\}\{failure\_test\_code\} \\
	&\text{Below is the test log}\\
	&\{error\_log\} \\
	&\text{Try to generate }\{file\_name\}\text{ again to pass the tests. RETURN CODE ONLY.}
\end{align*}

For all three prompt variants, we measure $pass@1$ ($n=1$) against all 1000 problems of the benchmark. Also in each experiment, we apply one prompt variant only, and compare it against the control test using the original prompt (Equation (\ref{eq:prompt})). Table \ref{tab:prompt_experiment} summarizes the relative performance gains/loss of each variant.
\begin{table}[h]
	\caption{Prompt Experiments: $Pass@1$ Gain/Loss}
	\centering
	\begin{tabular}{|l|c|c|c|}
		\hline
		& \textbf{System Prompt} & \textbf{Verbose Comment} & \textbf{Error Debugging} \\
		\hline
		gpt-4o & -1.3\% & -4\% & -56\% \\
		claude-3.5-sonnet & 6.3\% & -1\% & 38\% \\
		deepsseek-coder-v2-instruct & -18.2\% & 7.5\% & -79\% \\
		gemini-1.5-pro & 6.3\% & 2\% & 22\% \\
		llama-v3-70b-instruct & 8.5\% & -7.7\% & 111\% \\
		\hline
	\end{tabular}
	\label{tab:prompt_experiment}
\end{table}

To our surprise, we are unable to find a prompt variant delivering universally positive (or negative) impacts to all LLMs. Also we observe the huge swing in the error debugging column. The situation is unique here because this technique is not needed if the model output is correct on the first try. Strong models like GPT-4o can produce high $Pass@1$ ($n=1$) closed to 0.9, which significantly shrinks the sample size.

As such, we can not recommend LLM users to adopt or avoid any prompting technique we have experimented. More investigations are definitely needed here, especially why a seemingly harmless prompt variant should cause regression. 

Also for the sake of fair evaluation, we decide to use the original prompt (Equation (\ref{eq:prompt})) in our full-scale evaluation.
\subsection{Results}
Table \ref{tab:performance} shows results of full-scale evaluation in the form of leaderboard. Models are ranked by $Pass@10$. where gpt-4o is the winner. Results of $Pass@5$ completely agrees with this ranking. Results of $Pass@1$ also largely agrees, except claude-3.5-sonnet which surpasses gpt-4o as the winner.
\begin{table}[h]
	\centering
	\begin{tabular}{|l|c|c|c|c|}
		\hline
		& Number of Parameters & $Pass@1$ & $Pass@5$ & $Pass@10$ \\
		\hline
		gpt-4o & N/A & 0.8702 & 0.9013 & 0.909 \\
		claude-3.5-sonnet & N/A & 0.8808 & 0.8845 & 0.886 \\
		gpt-4o-mini & N/A & 0.8271 & 0.8534 & 0.858 \\
		deepseek-coder-v2-instruct & 236B & 0.7002 & 0.8087 & 0.827 \\
		gemini-1.5-pro & N/A & 0.6820 & 0.7678 & 0.795 \\
		gemini-1.5-flash & N/A & 0.5700 & 0.6427 & 0.663 \\
		deepseek-coder-v2-lite-instruct & 16B & 0.4610 & 0.6144 & 0.653 \\
		mixtral-8x22b-instruct & 176B & 0.3080 & 0.4821 & 0.533 \\
		llama-v3-70b-instruct & 70B & 0.3330 & 0.4462 & 0.489 \\
		mixtral-8x7b-instruct & 56B & 0.1270 & 0.1960 & 0.218 \\
		llama-v3-8b-instruct & 8B & 0.0679 & 0.1183 & 0.139 \\
		\hline
	\end{tabular}
	\caption{$pass@k$ over all LLMs}
	\label{tab:performance}
\end{table}

We believe the ranking of this leaderboard meets expectations of the LLM community. It is worth calling out the truly impressive performance of open source LLMs, especially Deepseek.

Also when comparing the large and small models of the same family, it is evident that scaling laws still holds. In fact, of the six open source models whose parameter counts are available, their performance ranking aligns with their size ranking, except deepseek-coder-v2-lite-instruct which itself is a coder model.

\section{Conclusion and Future Works}\label{sec:conclusion}
This report presents our preliminary contribution to LLM code generation. We proposed WebApp1K, a practical and comprehensive benchmark focusing on web app development. To recap findings via this benchmark:
\begin{enumerate}
	\item Open-source LLMs deliver impressive results
	\item Model size plays pivotal role to code generation performance
	\item No prompting techniques have been found to either universally help all LLMs or significantly help a single LLM
\end{enumerate}

We plan for the following future works.

First, more frontier LLMs are released as of the writing of this report, namely Llama 3.1\cite{llama3.1} and Mistral Large 2\cite{mistral-large-2}. We will report our evaluation results as soon as possible.

Second, we need to make the benchmark more difficult, as it is already saturated by top performers, i.e. GPT-4o and Claude 3.5. While this benchmark, as it stands now, can help other models to improve themselves, we need to make improvement space for the best models, as bigger and stronger models will be released sooner rather than later.

Finally, this work produce lots of error logs which we will dive deeper. We expect to uncover useful insights, which could help models of all rankings to continue the hill climbing.
\section{Appendix: Applications of WebApp1K}
\begin{table}[h]
	\caption{Applications of WebApp1K}
	\centering
	\begin{tabular}{|c|p{12cm}|}
		\hline
		\textbf{Name} & \textbf{Overview} \\
		\hline
		blogging & A content management system for creating and managing blogs, with features like user registration, post creation, categorization, commenting, and SEO optimization. \\
		\hline
		customer support & A help desk application where users can submit support tickets, track their status, access a knowledge base, and chat with support agents. \\
		\hline
		e-commerce & A fully functional e-commerce site with features like product listings, shopping cart, user authentication, order processing, and payment integration. \\
		\hline
		event management & An app for organizing events, including event creation, ticket sales, attendee registration, and scheduling \\
		\hline
		fitness tracking & An application for tracking fitness activities, setting goals, monitoring progress, and integrating with wearable devices. \\
		\hline
		inventory management & A web application designed to help businesses track and manage their inventory. Features include product cataloging, stock level monitoring, automated reorder alerts, supplier management, sales and purchase order processing, and detailed reporting on inventory performance. \\
		\hline
		job board & A job listing site where employers can post job openings and job seekers can search and apply for jobs. \\
		\hline
		music streaming & A platform for streaming music, creating playlists, and discovering new artists. \\
		\hline
		news aggregator & A news platform that aggregates articles from various sources, categorizes them, and allows users to customize their news feed. \\
		\hline
		online learning & An LMS where users can enroll in courses, watch videos, complete quizzes, track progress, and receive certificates. \\
		\hline
		online marketplace & A platform for buying and selling goods, similar to eBay, with features like user ratings, bidding, and secure transactions. \\
		\hline
		personal finance & A tool for managing personal finances, including expense tracking, budget planning, report generation, and financial goal setting. \\
		\hline
		pet care & a web application designed to help pet owners maintain a detailed record of their pet's health, activities, and milestones. \\
		\hline
		photo gallery & An application for uploading, organizing, and sharing photos, with features like tagging, album creation, and social sharing. \\
		\hline
		real estate & A platform for listing and searching real estate properties, with features like property details, image galleries, map integration, and contact forms. \\
		\hline
		recipe sharing & A platform where users can share, search, and save recipes, with features like ingredient lists, cooking instructions, and user ratings. \\
		\hline
		social media & A social media platform where users can create profiles, post updates, follow others, like and comment on posts, and manage a feed of updates. \\
		\hline
		task management & An application for managing tasks and projects, with features like task creation, assignment, progress tracking, and notifications. \\
		\hline
		travel planning & An app for planning and booking travel, including flight and hotel searches, itinerary creation, and travel recommendations \\
		\hline
		weather & An app that provides real-time weather updates, forecasts, and severe weather alerts. \\
		\hline
	\end{tabular}
	\label{tab:scenarios}
\end{table}
\bibliographystyle{plain}
\bibliography{ref}
\end{document}